\documentclass[aps,prx,preprint,superscriptaddress,longbibliography]{revtex4-1} 

\usepackage[utf8]{inputenc}
\usepackage{graphicx}  
\usepackage{dcolumn}   
\usepackage{bm}        
\usepackage{amssymb}   
\usepackage{amsmath}   
\usepackage{hyperref} 

\begin{document}

\title{Diversity improves performance in excitable networks}

\author{Leonardo L. Gollo}%
\affiliation{Systems Neuroscience Group, QIMR Berghofer Medical Research Institute, Brisbane, QLD 4006, 
Australia}

\author{Mauro Copelli}%
\affiliation{Departamento de F{\'\i}sica, Universidade Federal de Pernambuco, 50670-901 Recife, PE, Brazil}%

\author{James A. Roberts}%
 \affiliation{Systems Neuroscience Group, QIMR Berghofer Medical Research Institute, Brisbane, QLD 4006, 
Australia}

\begin{abstract}

As few real systems comprise indistinguishable units, diversity is a hallmark of nature.
Diversity among interacting units shapes properties of collective behavior such as synchronization and information transmission.
However, the benefits of diversity on information processing  at the edge of a phase transition, ordinarily assumed to emerge from identical elements, remain largely unexplored.
Analyzing a general model of excitable systems with heterogeneous excitability, 
we find that diversity can greatly enhance optimal performance (by two orders of magnitude) when distinguishing incoming inputs.
Heterogeneous systems possess a subset of specialized elements whose capability greatly exceeds that of the nonspecialized elements.
Nonetheless, the behavior of the whole network can outperform all subgroups. 
We also find that diversity can yield multiple percolation, with performance optimized at tricriticality.  
Our results are robust in specific and more realistic neuronal systems comprising a combination of excitatory and inhibitory units, 
and indicate that diversity-induced amplification can be harnessed by neuronal systems for evaluating stimulus intensities. 
\end{abstract}


\maketitle

\section*{Author Summary}

Diversity is ubiquitous in natural and artificial systems, primarily arising due to specialization. Specialized units have a clear role in optimally performing specific functions, but what is the role of the remaining non-specialized units? Surprisingly, we find that non-specialized units are fundamental for distinguishing the intensity of incoming inputs in excitable systems. Although non-specialized units themselves are inefficient at such intensity coding, their presence greatly enhances the performance of both the specialized units and the system as a whole. Our findings highlight the importance of combining both specialized and non-specialized units for optimal collective behavior, and indicate that diversity is more important than previously thought.

\section{Introduction}

In numerous physical~\cite{dagotto05}, biological~\cite{weng99} and social~\cite{silverberg13} systems, complex phenomena 
(including nonlinear computations~\cite{gollo09}) 
emerge from the interactions of many simple units.
Such interactions in a network of simple (linear-saturating-response) units 
generate nonlinear transformations that give rise to 
optimal intensity coding at criticality---the edge of a phase transition~\cite{kinouchi06, shew09, chialvo10}.
However, optimal collective responses often require diversity~\cite{tessone06}.
Clear examples of such optimization can be found in collective sports, business, and co-authorship 
in which different positions or roles require specific sets of skills contributing to the overall performance in their own way.

Diversity in the nervous system, for example, appears in morphological, 
electrophysiological, and molecular properties across neuron types and among neurons within a single type~\cite{sharpee14}, and also in the connectome~\cite{sporns11}, i.e., in how neurons and brain regions are connected. 
A large body of work has been devoted to show the role of heterogeneous connectivity and network topology 
in shaping the network dynamics~\cite{fornito15,misic15,gollo15,martin15,gollo14,restrepo14,villegas14,matias14,gollo14b,moretti13,larremore11,rubinov11,honey10,rubinov09,honey09,honey07}. 
In particular, for example, in the case of resonance-induced synchronization~\cite{gollo14}, the presence or not of a single backward connection 
may define whether synchronization or incoherent neural activity is expected in cortical motifs and networks, which has also been confirmed in a synfire chain configuration~\cite{moldakarimov15,claverol15}. 

Crucially, diversity in the intrinsic dynamic behavior of neurons is also fundamental and 
can shape general aspects of the network dynamics~\cite{vladimirski08,mejias12}. 
The role of the inherent diversity among nodes, which in many systems is at least as notable as the connectivity and network topology themselves, 
has comparatively remained largely unexplored.
In particular, although numerous recent works have focused on optimizing features of criticality for the different network topologies~\cite{haldeman05,kinouchi06,copelli07b,assis08,shew09, chialvo10,larremore11,shew11,yang12,mosqueiro13,moretti13,gollo13,haimovici13,plenz14}, for convenience identical units are ordinarily assumed and the role of nodal intrinsic diversity on the collective behavior thus remains unexplored.

\begin{figure}[!h]
\begin{center}
\includegraphics[angle=0,width=0.85\columnwidth]{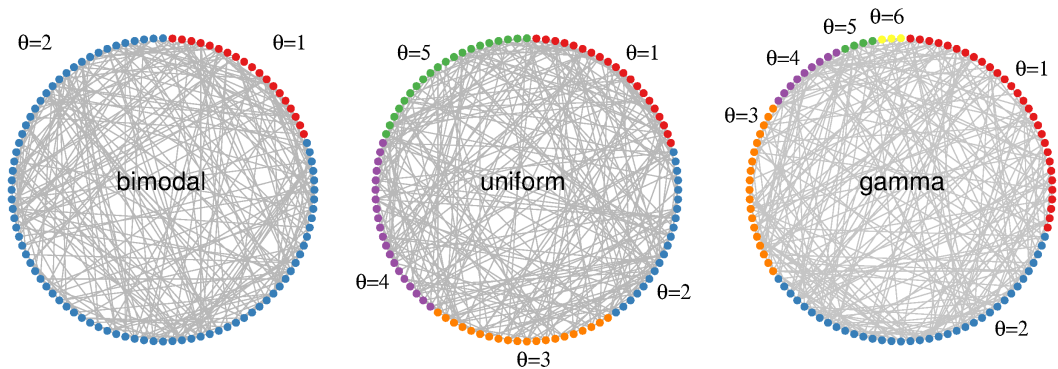}
\caption{\label{network} {\bf Threshold distributions in random networks.} 
  Threshold $\theta$ indicates the minimal number of coincident excitatory
  contributions required to excite a quiescent unit.
  Left panel, bimodal distribution with 80\% integrators ($\theta=2$).
  Middle panel, uniform distribution with $\theta_\mathrm{max}=5$.
  Right panel, gamma distribution with shape parameter $a=2$, and scale parameter $b=1$. 
  }
\end{center}
\end{figure}

Here we analyze the collective behavior at criticality in the presence of diversity in the excitability, which proves to be a crucial factor for the network performance. 
We show that the task of distinguishing the amount of external input, quantified by the dynamic range, 
can be substantially improved in the presence of heterogeneity.
The influence of non-specialized units improves performance by enhancing the capabilities of both the whole network and of specialized subpopulations.
We show the constructive effects of diversity in simple bimodal and uniform distributions, in more realistic gamma distributions (see Fig.~\ref{network}), and the robustness in networks combining excitatory and inhibitory units.

\section{Results}

\subsection*{Excitable networks with heterogeneous excitability}

Employing a general excitable model [susceptible-infected-refractory-susceptible (SIRS)], 
we characterize the dynamics and identify the constructive role of diversity in excitable networks and neuronal systems.
Node dynamics are given by cellular automata with discrete time and states 
[0 (quiescent or susceptible), 1 (active or infected), 2 (refractory)].
Synchronous update occurs at each time step (of 1 ms) obeying the rules: 
An active node $j$ becomes refractory with probability 1, a refractory node becomes quiescent with probability $\gamma=0.5$, 
and a quiescent node becomes active either by receiving external input (modeled by a Poisson process with rate $h$),  
or by receiving at least $\theta^j$ contributions from active neighbors each transmitted with a probability $\lambda$.
Diversity is introduced in the threshold variable $\theta^j$ of each node $j$ such that nodes with low threshold require fewer coincidental stimuli, 
being thus easily and more often excited by active neighbors than nodes with high threshold.
For concreteness, we used Erd\H {o}s-R\'{e}nyi random networks with size $N=5000$ and mean degree $K=50$, but
our results generalize to other sizes, connectivities, and topologies.

\subsection*{Mix of specialized and nonspecialized nodes outperforms either alone}

\begin{figure*}[!ht]
\begin{center}
\includegraphics[angle=0,width=1.0\columnwidth]{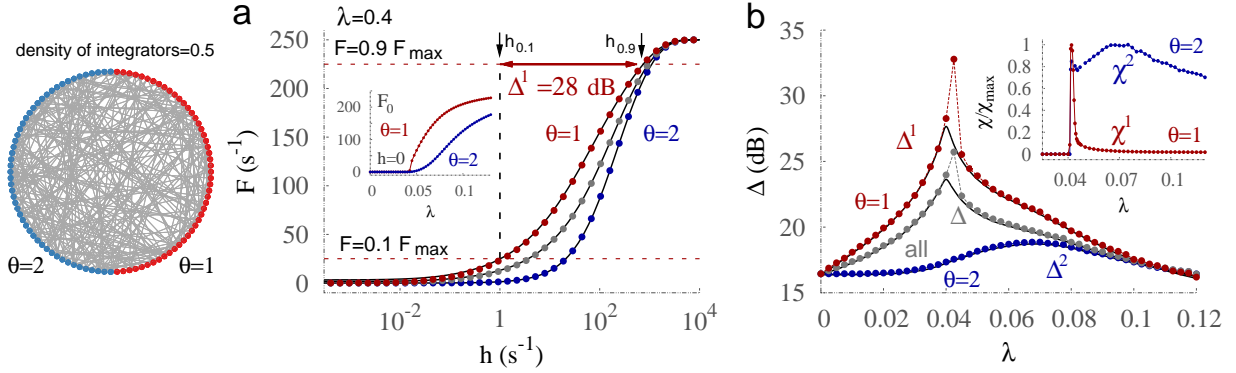}
\caption{\label{bimodal} {\bf A specialized subpopulation emerges with diversity.} 
 Bimodal distribution with equal numbers of integrators ($\theta=2$) and non-integrators ($\theta=1$). 
  {\bf a,} Response curves (mean firing rate $F$ versus stimulus rate $h$) for the subpopulations of $\theta=2$ (blue), $\theta=1$ (red), and the whole network (gray). Variables $F_{0.1}$ and $F_{0.9}$ (red dashed lines), and $h_{0.1}$ and $h_{0.9}$ (black arrows) are used to calculate the dynamic range $\Delta^1$ (red arrow) for the subpopulation with $\theta=1$, where $F_{x}=F_0+ x F_\mathrm{max}$, $h_{x}$ is the corresponding input rate to the system, and $F_0$ is the firing rate in the absence of input. 
    Solid black lines correspond to the mean-field approximation (see Methods). 
    Inset: Spontaneous activity $F_0$ versus coupling strength $\lambda$. 
  {\bf b,} Dynamic range $\Delta$ is optimized for different coupling strengths $\lambda$ for the two subpopulations.
  Inset: {\emph {Susceptibility}} $\chi^{\theta}$ for the two corresponding subpopulations; 
  susceptibility maxima coincide with the peaks of the dynamic range. 
    Susceptibility is operationally defined in the Methods. 
  }
\end{center}
\end{figure*}

Our analysis focuses on the input-output response function of networks subjected to external driving $h$ whose intensity varies over several orders of magnitude, as is commonly observed in sensory systems, for example.
As depicted in Fig.~\ref{bimodal}a, response functions ($F$) 
exhibit a sigmoidal shape with lower output rates (defined as the mean activity of the network or a subset thereof) for weak stimuli and high rates for strong stimuli.
In this simple case diversity is introduced by a discrete bimodal distribution (see Fig. 1), where 
half the units are so-called integrators with $\theta=2$, and the other half are nonintegrators with $\theta=1$.
From the shape of the response functions we quantify the range in which the amount of input can be coded by the output rate (Fig.~\ref{bimodal}a).
This dynamic range $\Delta=10 \log_{10} (h_{0.9}/h_{0.1})$ is a standard measure that neglects the confounding ranges of too small sensitivity [top  10\% ($F>F_{0.9}$) and bottom  10\% ($F<F_{0.1}$)], and quantifies how many decades of input $h$ can be reliably coded by the output activation rate $F$ (see caption of Fig.~\ref{bimodal}a for further details).

Although isolated units ($\lambda=0$) code input intensity very poorly (small $\Delta$), 
increasing the contribution from neighbors (by increasing the transmission probability~$\lambda$) substantially enhances the dynamic range (Figs.~\ref{bimodal}b and~\ref{bimodalImprovement}).
However, this occurs only for coupling smaller than a critical value $\lambda_c$, at which a phase transition to self-sustained activity occurs (e.g., insets of Fig.~\ref{bimodal}a and Fig.~\ref{Gbimodal}a).
As the coupling strength increases beyond the critical value, the dynamic range decays because the effective output range is reduced by increasing levels of self-sustained activity~\citep{kinouchi06}.
In this simple bimodal case the phase transition occurs at different $\lambda$ values for the two subpopulations, evidenced by 
both dynamic range $\Delta^{\theta}$ and {\emph {susceptibility} $\chi^{\theta}$ (Fig.~\ref{bimodal}b and its inset).
The critical value of the coupling (curve's peak) is larger for integrators than for nonintegrators.
Moreover, as evidenced by the difference between the maximum dynamic range of each subpopulation ($\Delta^1_\mathrm{max}-\Delta^2_\mathrm{max}\simeq15$~dB), nonintegrators greatly outperform integrators.

\begin{figure*}[!ht]
\begin{center}
\includegraphics[angle=0,width=0.4\columnwidth]{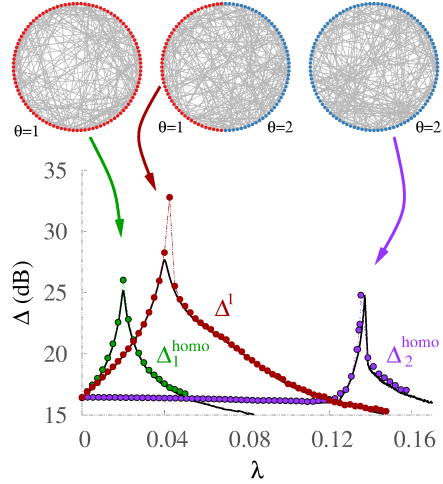}
\caption{\label{bimodalImprovement} {\bf Threshold diversity improves performance.} 
Comparison of dynamic range for networks with homogeneous thresholds with $\theta=1$ (green, $\Delta^\mathrm{homo}_1$) and $\theta=2$ (purple, $\Delta^\mathrm{homo}_2$) with
   the $\theta=1$ subpopulation of the bimodal distribution (red,  $\Delta^1$). Solid black lines correspond to the mean-field approximation (see Methods). 
  }
\end{center}
\end{figure*}

In the presence of diversity the  {\emph {specialized}} subpopulation of nonintegrators ($\Delta^1$) outperforms the two extreme cases with no diversity (homogeneous distribution) in which all units are either integrators $\Delta^\mathrm{homo}_2$
or nonintegrators  $\Delta^\mathrm{homo}_1$ (Fig.~\ref{bimodalImprovement}).
This happens because the response of the specialized units 
improves when they can also take advantage of the contribution of the other 
subpopulation of integrators, which require simultaneous neighboring stimulation to be effective. 
In the presence of integrators the network becomes more disconnected, requiring stronger coupling to switch to the active state.
Therefore, due to a stronger coupling, the amplification of weak stimuli at criticality and thus the dynamic range are greater than in the absence of diversity. 
Thus, the presence of {\emph {prudent}} units delays the critical transition and provides {\emph {gullible}} units additional sensitivity to distinguish stimulus intensity.
Remarkably, however, having all nodes behave like the specialized ones impairs performance.

\subsection*{Tricriticality optimizes coding performance}

\begin{figure*}[!h]
\begin{center}
\includegraphics[angle=0,width=1.0\columnwidth]{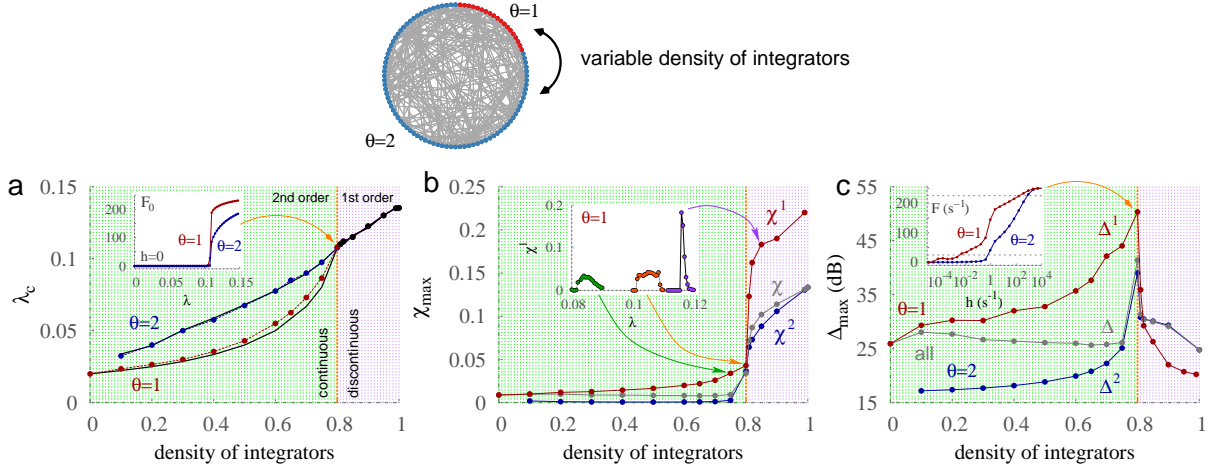}
\caption{\label{Gbimodal} {\bf Performance is optimized at tricriticality with a critical density of integrators and a critical coupling strength.} 
   General bimodal distribution with varying densities of integrators ($\theta=2$) and non-integrators ($\theta=1$). 
      {\bf a,} Critical coupling strength ($\lambda_c$) as a function of the density of integrators for the two subpopulations.
   Curves collide at a tricritical point (orange line), separating regimes with continuous (2nd order, green) and discontinuous (1st order, purple) phase transitions.
    Inset: Spontaneous activity $F_0$ versus coupling strength $\lambda$ for the critical density of integrators.
   {\bf b,} Maximum susceptibility $\chi_\mathrm{max}$ versus density of integrators.
   Inset: Susceptibility of subpopulation with $\theta=1$ versus coupling strength for three integrator densities (0.75, 0.8, 0.85).
   {\bf c,} Maximum dynamic range $\Delta_\mathrm{max}$ versus density of integrators. 
    Inset: Response curves at the tricritical point ($\lambda=0.1075$). 
  }
\end{center}
\end{figure*}

Henceforth, since criticality optimizes performance, we focus on characterizing the critical behavior for various types of diversity in the excitability.
Varying the density of integrator units (with $\theta=2$) while the rest are nonintegrators, we find a critical point separating two regimes (Fig.~\ref{Gbimodal}a): 
For a low density of integrators (green region) the phase transition to the regime of spontaneous activity is continuous (transcritical bifurcation in the mean-field equations for the model, see  Methods); 
for a high density of integrators (purple region) the phase transition to the regime of spontaneous activity is discontinuous (saddle-node bifurcation in the mean-field equations)~\citep{gollo12b}.
The critical coupling ($\lambda_c$) grows with the density of integrators for both the subpopulation of integrators (blue) and nonintegrators (red) and these curves collapse at the tricritical point (orange line).
Apart from this collapsing of critical-coupling curves, 
the maximum susceptibility also changes qualitatively at the transition between the regions undergoing continuous and discontinuous phase transitions (Fig.~\ref{Gbimodal}b). 
Strikingly, optimal performance occurs at this transition (Fig.~\ref{Gbimodal}c):  
The maximum dynamic range for generalized bimodal distributions occurs at the {\emph {tricritical} point where 
the sensitivity is more than two orders of magnitude larger than in the absence of diversity ($\Delta_1^\mathrm{homo}$ in Fig.~\ref{bimodalImprovement}).

\subsection*{Diversity can yield multiple percolation}

Large dynamic ranges also occur at criticality in other distributions such as the uniform distribution.
In this case, the number of units with threshold $\theta$ is evenly distributed between 1 and $\theta_\mathrm{max}$, as depicted in the middle panel of Fig~\ref{network} 
for an exemplar case with $\theta_\mathrm{max}=5$. 
Notably, for the uniform distribution, $\Delta^1_\mathrm{max}$ is much greater than $\Delta_\mathrm{max}$ of the other subpopulations (Fig.~\ref{uni}a) and of the whole network (inset).

\begin{figure*}[!h]
\begin{center}
\includegraphics[angle=0,width=1.0\columnwidth]{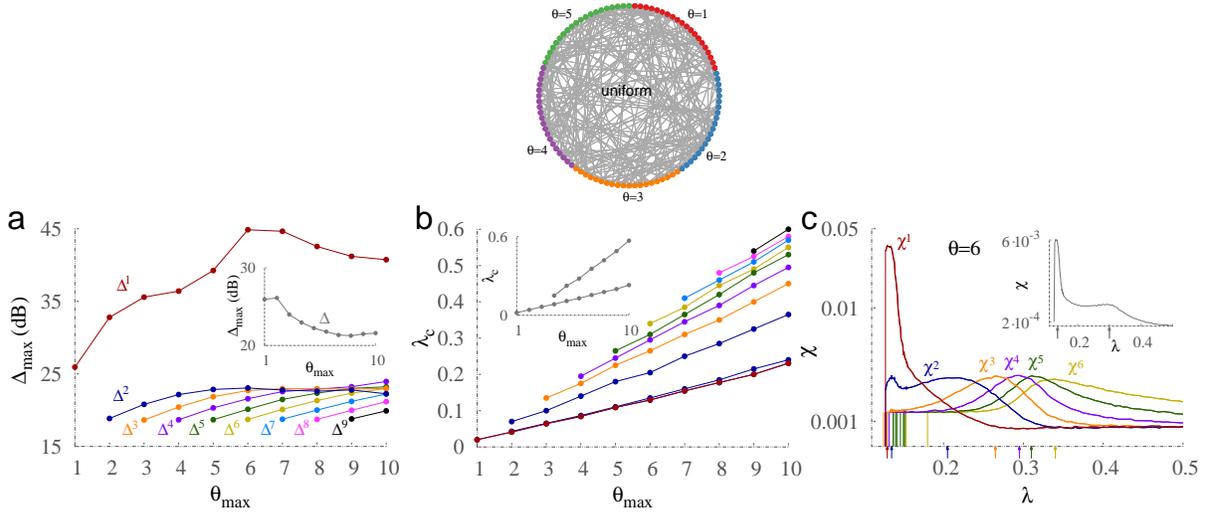}
\caption{\label{uni} {\bf Multiple percolation and optimal performance in uniform distributions of thresholds.} 
  {\bf a,} Maximum dynamic range $\Delta_\mathrm{max}$ versus the maximum threshold of the uniform distribution $\theta_\mathrm{max}$ for each subpopulation, 
  and the whole network (inset).
 {\bf b,} Critical coupling strength ($\lambda_c$) as a function of $\theta_\mathrm{max}$ for each subpopulation. 
 The whole network (inset) exhibits two peaks for  $\theta_\mathrm{max}>3$.
  {\bf c,} Susceptibility versus coupling strength for each subpopulation, and the whole network (inset).
  Arrows at the bottom of the panel identify the critical couplings. 
  }
\end{center}
\end{figure*}

In contrast to the bimodal distribution (Fig.~\ref{Gbimodal}a), the critical coupling curves of the subpopulations for the uniform distribution grow with $\theta_\mathrm{max}$  {\emph {without}} collapsing (Fig.~\ref{uni}b).
Hence, the system exhibits multiple critical couplings.
However, the network taken as a whole exhibits only two peaks of susceptibility (insets of Figs.~\ref{uni}b,c):
the lowest curve matches the value of the subpopulation with  $\theta=1$, and the other corresponds to an average of all subpopulations.
Figure~\ref{uni}c displays the curves of the susceptibility for each subpopulation and the whole network (inset).
The larger the $\theta$ of the subpopulation, the greater the coupling required to optimize the susceptibility, 
leading to a subpopulation hierarchy.

\subsection*{More realistic scenarios}

\subsubsection{Gamma distribution}

\begin{figure*}[!b]
\begin{center}
\includegraphics[angle=0,width=0.8\columnwidth]{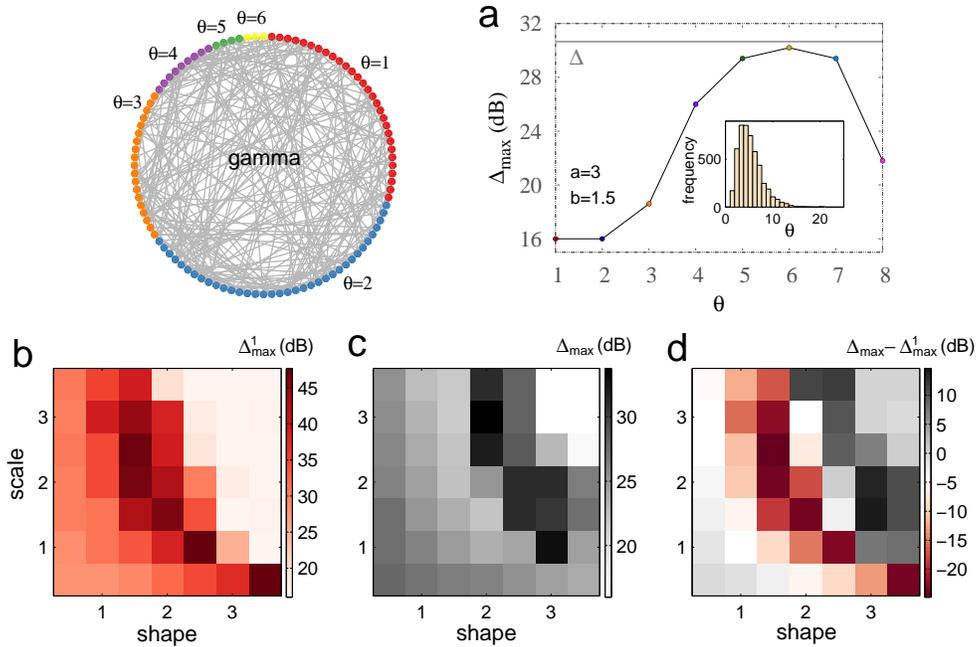}
\caption{\label{gamma} {\bf Optimal performance in gamma distributions of thresholds: the whole can outperform any of its parts.} 
  {\bf a,} Maximum dynamic range for various subpopulations and the whole network (solid gray line).
  Inset: Gamma distribution of threshold values for shape parameter $a=3$, and scale parameter $b=1.5$. 
  {\bf b-d,} Maximum dynamic range versus scale and shape parameters of the gamma distribution.
  {\bf b,} Specialized (sensitive) subnetwork; 
  {\bf c,} the whole network;  
  {\bf d,}  difference between the whole network and the specialized subnetwork.
  }
\end{center}
\end{figure*}

The gamma distribution is more general and presumably more realistic than the bimodal and uniform distributions.
As presented in the Methods and illustrated in Fig~\ref{network}, 
it is described by two independent parameters, shape {\emph {a}} and scale factor {\emph {b}}, 
and generalizes the exponential, chi-squared, and Erlang distributions.
Exploring random networks with thresholds given by discrete gamma distributions 
(see an exemplar case in the inset of Fig.~\ref{gamma}a), 
we find large dynamic ranges (Figs.~\ref{gamma}a-d). 
The maximum dynamic range for both the subpopulation with $\theta=1$ and the whole network can reach $\sim 40$~dB.
Moreover, as shown in Fig.~\ref{gamma}a, the dynamic range for the whole network can outperform all subpopulations.

\subsubsection{Networks with excitatory and inhibitory nodes}

\begin{figure}[!b]
\begin{center}
\includegraphics[angle=0,width=0.8\columnwidth]{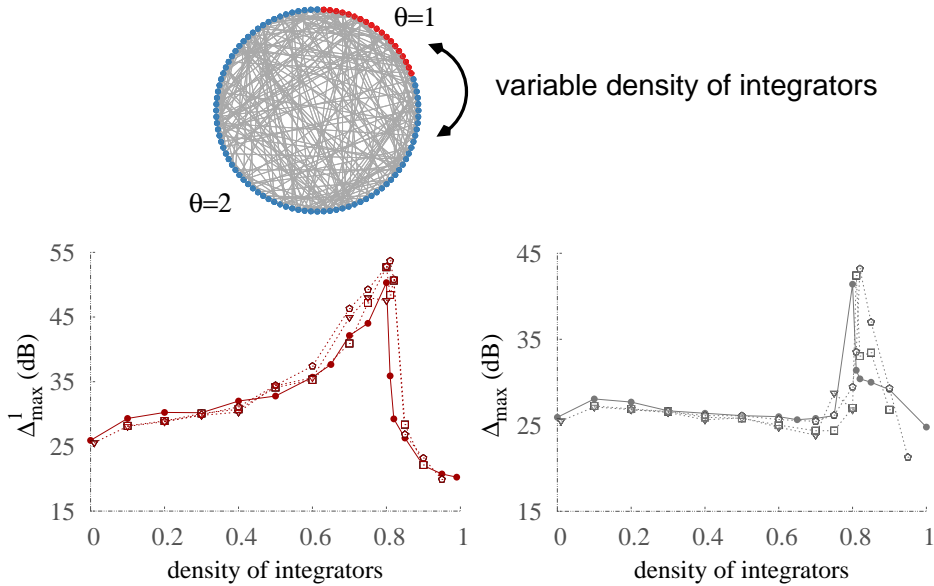}
\caption{\label{inhibition} {\bf Robustness of optimization in a network with 20\% inhibitory units.}
  Thresholds are drawn from a bimodal distribution of integrators ($\theta=2$) and non-integrators ($\theta=1$). 
  Maximum dynamic range versus coupling strength for the specialized subnetwork  (left) and the whole network (right) for different types of inhibitory units: 
    nonintegrating (triangles),  
    integrating (pentagons), 
    half integrating and half nonintegrating (squares), 
  and the case without inhibition (filled circles). 
  }
\end{center}
\end{figure}

Our main result that performance can be substantially enhanced with diversity is also robust with respect to the presence of inhibition.
Inhibition has two effects on the response function, influencing the dynamic range in opposite ways.
On the one hand, inhibition prevents a rapid increase in the firing rate for small input.
On the other hand, it prevents saturation for large input.
The first effect tends to reduce the dynamic range whereas the second effect tends to increase it.
In the absence of diversity, the overall effect reported is a small reduction in the network dynamic range~\cite{pei12}.
In the presence of diversity, however, we find the overall effect counterbalanced and inhibition does not alter the enhancement of $\Delta$.
Here we assume that the distribution of $\theta$ is bimodal and
20\% of the units (neurons) are inhibitory.
After an inhibitory neuron spikes, post-synaptic quiescent neurons receive inhibition with probability $\lambda$. 
Upon arrival, inhibition prevents the unit from spiking within a time-step period irrespective of the number of excitatory active neighbors.
Figure~\ref{inhibition} shows the robustness of the maximum dynamic range 
in the presence of inhibition.
Regardless of whether the inhibitory units are integrators (pentagons), nonintegrators (triangles), or a mix of both (square) the dynamic ranges 
are very similar to the case without inhibition (filled circles).
Although inhibition has been shown to crucially shape the network dynamics~\cite{larremore14}, and diversity in excitatory and inhibitory populations may have different effects~\cite{mejias14}, 
we found that in the presence of diversity inhibition produces only minimal quantitative differences in the coding performance of networks.

\section{Discussion}

Diversity has been a keystone of the recruitment theory~\cite{cleland99} that proposed the first explanation for how animals can distinguish incoming input spanning many orders of magnitude, even when each individual sensory neuron distinguishes only a narrow dynamic range. 
The proposed mechanism there requires many neurons exhibiting responses tuned to specific (short) ranges of input but with the ensemble of specific ranges spanning  several orders of magnitude. 
However, to satisfy this criterion sensory neurons would need to have a density of receptors also varying across orders of magnitude, which is not found experimentally~\cite{chen94,cleland99}.  
A competing hypothesis claims that diversity is not required, but instead nonlinear interactions are sufficient for sensory systems to cope with incoming input varying over many orders of magnitude~\cite{kinouchi06,copelli07}. 
Remarkably, our revisited version of the recruitment theory reconciles the two proposals employing the key ingredient of each one:  
mutual (non-linear) interactions, which amplify the dynamic range of isolated neurons, and 
intrinsic diversity in the excitability, which requires small variability (and not variations of orders of magnitude as in the previous version).
Therefore, showing that diversity enhances the dynamic range of response functions, 
we establish a revisited recruitment theory with solid grounds.

Although we have focused on a specific task of distinguishing stimuli intensity, sensory systems also need to handle various other features. 
As a byproduct and another advantage of diversity, nonspecialized units may execute and specialize in other functions.
For example, as recently reported in the moth olfactory system~\cite{rospars14}, a concurrent function of the detection of stimulus intensity is the ability to respond promptly to external stimuli. 
Under evolutionary pressure, the ability to execute such complementary functions likely takes advantage of diversity to improve its own performance.

We have demonstrated the benefits of diversity at criticality for different simple distributions of excitability (as requested in the recent literature~\cite{baroni14}). 
Furthermore, for the first time we provide evidence that the well-known advantages of criticality are magnified at tricriticality. 
The optimal performance in the simple case of two type of units is found at a tricritical point with a critical coupling separating the active/inactive phases and a critical density of integrators separating the regimes of continuous/discontinuous phase transitions. 
Even though a continuous phase transition has been proposed for the brain~\cite{chialvo10}, hysteresis and metastability observed in models~\cite{gollo12b,wilson72} and experiments~\cite{kastner14} suggest that discontinuous phase transitions may also play a functional role.

The dynamics of excitable networks exhibits two regimes: percolating (active phase) and non-percolating (inactive phase).
As recently shown~\cite{colomer14}, percolation in core-periphery networks with sufficient clustering leads to double percolation, 
in which core nodes percolate earlier than peripheral nodes.
Analogously, for bimodal distributions we found double percolation with the most excitable nodes activating for weaker coupling 
than integrators.
Moreover, we extended this phenomenon to arbitrarily high-order {\emph {multiple percolation}}, with subpopulation thresholds following a hierarchy of excitability.

\paragraph*{ Conclusion.} Minimal models play a key role to elucidate rich emergent dynamics that remain elusive. 
Following this approach and investigating the impact of diversity in the intrinsic excitability, we have shown that: 
(i) Diversity offers clear-cut advantages in distinguishing input with respect to homogeneous networks;  
(ii) At the tricritical point the system benefits from multiple critical instabilities, thereby optimizing performance; 
(iii) Subpopulations percolate in order of decreasing excitability;  
(iv) The collective response from the entire network can outperform all subpopulations; 
(v) The main results are robust to more realistic distributions, 
and can be applied to cortical systems composed of excitatory and inhibitory neurons.

\section{Methods}

\subsection{\bf Network Response} 
The initial condition for computing the firing rate corresponds to the active state. Nodes receive a strong input ($h=200$ Hz) for 0.5 s, followed by a transient period of 0.5 s with the corresponding input level ($h$) before computing the average firing rate of each subpopulation over a period of 5 seconds. The reported firing rate corresponds to the average over 5 trials.  

\subsection{\bf Mean-Field Approximation}
In the presence of diversity the mean field map is given by a set of equations for each subpopulation, exhibiting a particular sensitivity to neighboring signaling~\cite{gollo12b}.
For each subpopulation with threshold ${\theta_i}$, the density of refractory units  $R^{\theta_i}$ at time $t+1$ is given by $R_{t+1}^{\theta_i}=F_t^{\theta_i}+(1-\gamma)R_t^{\theta_i}$, where $F_t^{\theta_i}$ denotes the density of active units. 
The evolution of the density of active units
follows $F_t^{\theta_i}=Q_t^{\theta_i}[1-(1-h)(1-{\Lambda_t}^{\theta_i})]$, 
where $Q_t^{\theta_i}$ is the density of quiescent units, and 
${\Lambda_t}^{\theta_i}= \sum_{i=0}^{\theta_i-1} \binom{K}{i} (\lambda F_t)^i (1-\lambda F_t)^{K-i}$ 
represents the probability of not receiving at least $\theta_i$ neighboring contributions at time $t$, 
where $F_t$ is the weighted average of the density $d^{\theta_i}$ of active units in each subpopulation $F_t=\sum_{\theta_i} d^{\theta_i} F_t^{\theta_i}$.
Integrating this map~\cite{gollo12}, we find the stationary distributions ($F^{\theta_i}$) for each subpopulation. 

\subsection{\bf Susceptibility}
Here, susceptibility is operationally defined as $\chi^{\theta_i} =  \left \langle {\rho^{\theta_i}}^2 \right \rangle/  \langle {\rho^{\theta_i}} \rangle - \langle {\rho^{\theta_i}} \rangle$, where $\rho^{\theta_i}=F^{\theta_i}(h=0)$. It 
was calculated over 500 trials of 100 ms after transients of 0.5 s.

\subsection{\bf Gamma Distribution}
The discrete gamma distribution of thresholds is given by the smallest following integers drawn from the probability density function
$f(\theta)= \theta^{a-1} {e^{-\theta/b}} (b^a \Gamma(a))^{-1} $, where $a$ and $b$ are shape and scale parameters, respectively.

\bibliography{./Diversity}

\end{document}